Guerrilla Science: creating scientific instruments in "high tropicallity" conditions[1]

E. Altshuler
Group of Complex Systems and Statistical Physics, Physics Faculty, University of Havana, 10400 Havana, Cuba

*The 1980's was a flourishing time for Cuban physics, with various achievements ranging from the design of several experiments to be performed by a Cuban cosmonaut in 1980, to the synthesis of the first superconductor with critical temperature above 77 K shortly after being originally reported by US scientists. By the early 1990's, there was a profound economic crisis in the country. The situation strongly affected the availability of scientific instruments to Cuban physicists –a process that continues today. Doing science in such "High Tropicallity" scenario has been challenging. Many Cuban researchers have been forced to create competitive instruments with minimal resources. Here we put examples of some of these experimental setups, created by members of the Group of Complex Systems and Statistical Physics, University of Habana. Most of them involve the recycling of old laboratory instruments, and even the use of daily life devices.*

The sudden interruption of liquid nitrogen supply for performing superconductivity measurements, the necessity to stop air conditioning in laboratories for saving electricity, the impossibility to renew decaying equipment: these, and many other examples, illustrate what I call "Doing science in high tropicallity conditions" (HTC). The scenario is profusely illustrated in the book "Guerrilla Science: survival strategies of a Cuban physicist" [1]. In spite of the evident limitations to make cutting-edge science associated to HTC, I believe that a careful and open-minded selection of research topics can overcome the handicap to a substantial degree.

The discovery of High temperature superconductors by the end of the 1980's constituted an authentic scientific revolution reaching the University of Havana during a time a scientific maturity, especially in the field of ceramic materials. It is then not strange that professors and technicians of the Physics Faculty produced the first sample of the YBCO superconducting system a few months later that the first report published by C. W. Chu's group in the States. That early result produced a rapid development in Havana over the next few years. While the government helped with financial support for new equipment, we rapidly created ad-hoc devices like low temperature inserts by recycling old equipment, and even using daily-life materials. That kind of strategy allowed transport measurements in ceramic superconductors which constituted a full line of research for several years [2-8]. One of my favorite examples is the use of a copper-alloy Canadian coin to close one extreme of a stainless steel tube where a superconducting sample would be inserted, so it was insulated from a liquid nitrogen bath (somehow, we managed to weld the coin to the tube using conventional tin solder). Such "shortcuts" are sometimes surprisingly effective even in low-tropicallity conditions: a former student that wrote his Diploma under my supervision moved to a foreign country, and started his PhD in a state-of-the-art laboratory. After fighting with commercial, sophisticated equipment for more than a year without being able to obtain substantial results, he decided to reproduce in detail one of the low-temperature inserts we had made in Havana. It resulted in a dramatic acceleration of is PhD research. However, I must accept that some of my more sophisticated research –like the detection of superconducting vortex avalanches– critically required the collaboration of US and Israel groups [9,10].

Facing the 1990's crisis, we understood that a new strategy was needed in order to produce state-of-the-art science, in spite of severe material limitations. Sandpiles had been used as a paradigm of vortex avalanches. So, why not studying sandpile avalanches themselves? Or, why not studying the dynamics of granular materials, in general? Granular matter was becoming a very popular research subject by the end of the XX century. Starting by the year 2000, we accelerated the experimentation with true sandpiles. In order to construct an appropriate set up, we used all kinds of recycled equipment and materials: from Chinese meccano-toy parts, to a digital scale with a broken liquid-crystal display. The idea was to quantify the avalanche size distribution in a pile of beads slowly fed from above, and correlate it with the structural changes in the pile. The research was performed 100% in Cuba, and published in *Physical Review Letters* [11]. It was substantially enriched by a former Cuban student during his PhD at the University of Oslo. He used state-of-the-art equipment, but applied almost exactly the same experimental configuration we had designed in Cuba [12,13]. Our experience at home indicated that the use of parts from old equipment can sometimes accelerate the setting up of an experiment, since there is no need to wait for special materials or specific fabrication protocols. It decreases quite a bit the paperwork, too!

---
[1] Paper based on the oral contribution presented at the "XXXVIII Scientific Instrument Symposium" (Havana, 23-27 September 2019).

As we tried to set up the avalanche device in an inexpensive way, we attempted experiments with Cuban beach sands. They showed a surprising behavior that did not fit our "avalanches of all sizes" paradigm: the so-called "revolving rivers". After a few years and very inexpensive experiments, we wrote a number of papers in this matter [14,15,16]. The revolving rivers research illustrates a second strategy to cope with HTC: being proactive when unexpected phenomena appears –even if they contradict our expectations.

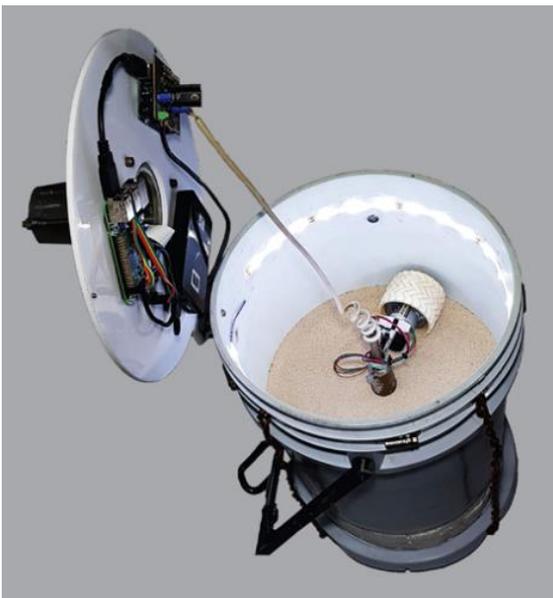

Fig. 1 Wheel-in-a-bucket. As the bucket falls at a controlled acceleration, the scaled-down Mars Rover wheel rolls on top of the sandy surface, and is filmed from top.

More recently, we have become interested in the penetration of a solid intruder into granular matter. One possibility is studying the penetration when the granular bed is fluidized due to lateral shaking. However, if the intruder sinks completely, it cannot be followed by using a camera. We proposed a technique called "lock in accelerometry" to follow the penetration dynamics without the use of cameras [17]. For that experiment, besides inexpensive wireless accelerometers, we used an electromagnetic shaker directly recovered from the university dump area.

A more spectacular piece of research involved a 15-meter tall Atwood machine used to study the penetration of an intruder into granular matter while it "feels" different effective gravitational accelerations –as if the experiment was performed on the surface of Mars, for example. The machine included a large pulley extracted from the external heat-exchange unit of large (broken) air conditioner; a recycled paint bucket was redefined as a "falling laboratory" attached to one of the sides of the Atwood machine, while the counter-weight made from a heavy air filter was attached to the other end. The bucket was filled with a light granular material and, while it fell, a ping-pong ball equipped with a wireless accelerometer inside would fall into the granular material, "feeling" an effective gravity different from Earth's. The ping-pong ball was released using the mechanisms of a discarded CD-player. It has resulted in a quite unique piece of research, that may shed light on why Mars rover wheels can be trapped into soft Martian sand [18]. Later on, we have used the same general idea to study the rolling efficiency of a small, 3D-printed Mars Rover wheel on sand [19,20].

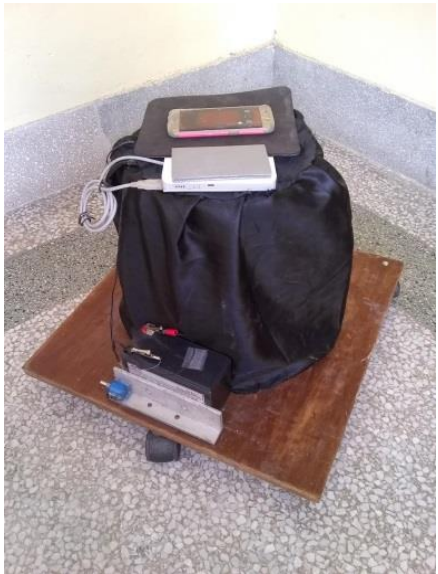

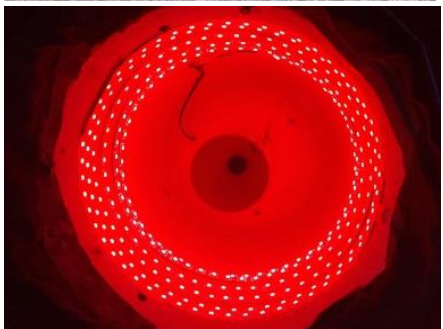

Fig. 2 Ant-in-a-bucket. Top: external view of the device. Bottom: Inner view of the bucket, looking up from the bottom.

After breaking the "phychological barrier" of partially quitting research in Solid State Physics (the most productive research line in Cuban Physics for decades), we were opened to all sorts of ideas. For example, assuming ants as a special kind of "active granular matter", we designed several experiments in ant dynamics, where a new technique was developed involving the use of a commercially available insect repellent [21-24]. Curiously, in order to use the relatively low resolution cameras at hand, we have specialized in one kind of ant: the *Atta insularis*, endemic of Cuba. It has two important properties for experimentation in High Tropicallity Conditions: they are relatively large, and they are black, so low-quality cameras yield images good enough for further analysis.

As we have seen, the "High Tropicallity Conditions" frame of mind can be eventually exported to labs in the western world, resulting in what we might call "Medium Tropicallity Conditions". I will end this short account by illustrating one of the most extreme cases dealing with this scenario. During a two-month trip to the University of Navarra (Spain) I was asked to introduce some experimental idea beyond their usual (and very strong!) research subject: the physics of discharging granular silos. I proposed that a "granular Maxwell daemon" could be created by introducing an asymmetric, mobile cage with an open door into a vibrating granular bed: I expected the grains to be "swallowed" by the cage due to the dissipative nature of the interactions between grains, and between grains and wall. But –at least during the first two weeks of my visit– the idea just did not work. However, while playing with the state-of-the-art vibrator I was using for the experiments, somehow I decided to create a new type of granular particle: I removed three rubber filaments from a broomstick, attached them to the rim of a Coca cola bottle cap with a certain inclination, and put the resulting device to vibrate vertically, looking like a three-legged spider. The new type of grain converted the vertical vibrational motion into quite smooth rotation around a vertical axis, so I called it *Vibrot*. It was hard to find an appropriate venue to publish such "off-mainstream" piece of research, but we finally found the right place [25] (The explanation of why vibrots work soonly came afterwards [26]). Months later, I would mention to my German colleague Thorsten Poschel that vibrots could be seen as just the beginning of a wider research subject: granular gases for which rotational degrees of freedom were very relevant to the dynamics. Thorsten sonly applied digital fabrication tools to mass-produce vibrots [27], and eventually started publishing a series of seminal papers in the matter [28].

Even when doing science in High Tropicallity Conditions may be fruitful and even fun, I should not recommend that work style for all scientific endeavors: some research areas –like vaccine design, for example– simply cannot afford starying at the margins of state-of-the-art technology. However, low-cost and rapidly spreading technologies like digital fabrication look like a reasonable bridge between High and Medium Tropically scenarios, at least in technical terms. However, a solid economy and stability are mandatory for doing science in Low Tropicallity Conditions…even in the case of tropical countries.

**References**

[1] E. Altshuler "Guerrilla Science, survival strategies of a Cuban Physicist" (Springer, 2017)
[2] E. Altshuler, S. García y A. Aguilar "Hysteretic Critical Currents in Y-Ba-Cu-O Superconductors: a Microstructural Approach". *Physica Status Solidi (a)* **120**: K169 (1990)
[3] E. Altshuler, S. García y J. Barroso "Flux Trapping in Transport Measurements of $YBa_2Cu_3O_{7-x}$ Superconductors: A Fingerprint of Intragrain Properties". *Physica C* **177**: 61 (1991)
[4] E. Altshuler, J. Musa, J. Barroso, A. R. R. Papa y V. Venegas "Generation of Hysteresis Jc(He) Curves in Ceramic YBa2Cu3O7-x Superconductors". *Cryogenics* **33**: 308 (1993)
[5] P. Muné, E. Altshuler y J. Musa "On the negative values of the geometric factors in the intragranular flux-trapping model and the hysteresis in the $J_c$ $(B_a)$ dependence of polycrystalline superconductors". *Physica C* **246**: 55 (1995)
[6] E. Altshuler, R. Cobas, A. J. Batista-Leyva, C. Noda, L. E. Flores, C. Martínez y M. T. D. Orlando "Relaxation of the transport critical current in high-$T_c$ polycrystals" . *Phys. Rev. B*. **60**: 3673 (1999)
[7] A. J. Batista-Leyva, R. Cibas, M. T. D. Orlando, C. Noda y E. Altshuler "Magnetic hysteresis of the zero-resistance critical temperature in YBaCuO, BiSrCaCuO and HgBaCaCuO superconducting polycrystals". *Physica C* **314**: 73 (1999)
[8] A. J. Batista-Leyva, R. Cobas, M. T. D. Orlando y E. Altshuler "Hysteresis and relaxation in $TlBa_2Ca_2Cu_3O_y$ superconducting polycrystals". *Supercon. Sci. Technol*. **16**: 857 (2003)
[9] E. Altshuler, T. H. Johansen, Y. Paltiel, Peng Jin, K. E. Bassler, O. Ramos, Q. Chen, G. F. Reiter, E. Zeldov y C. W. Chu "Vortex avalanches with robust statistics observed in superconducting niobium" *Phys. Rev. B* **70**: 140505 (R) (2004)
[10] E. Altshuler y T. H. Johansen "Experiments in vortex avalanches". *Rev. Mod. Phys*. **76**: 471 (2004)
[11] E. Altshuler, O. Ramos, E. Martínez, L. E. Flores y C. Noda "Avalanches in one-dimensional piles with different types of bases". *Phys. Rev. Lett*. **86**: 5490 (2001)
[12] O. Ramos, E. Altshuler y K. J. Måløy "Quasiperiodic events in an earthquake model". *Phys. Rev. Lett*. **96**: 098501 (2006)
[13] O. Ramos, E. Altshuler y K. J. Måløy "Avalanche prediction in a self-organized pile of beads". *Phys. Rev. Lett*. **102**: 078701 (2009)
[14] E. Altshuler, O. Ramos, E. Martínez, A. J. Batista-Leyva, A. Rivera y K. E. Bassler "Sandpile formation by revolving rivers" *Phys. Rev. Lett*. **91**: 014501 (2003)
[15] E. Martínez, C. Pérez-Penichet, O. Sotolongo-Costa, O. Ramos, K. J. Måløy, S. Douady y E. Altshuler "Uphill solitary waves in granular flows". *Phys. Rev. E* **75**: 031303 (2007)
[16] E. Altshuler, R. Toussaint, E. Martínez, O. Sotolongo-Costa, J, Schmittbuhl y K. J. Måløy "Revolving rivers in sandpiles: from continuous to intermittent flows". *Phys. Rev. E* **77**: 031305 (2008)

[17] G. Sánchez-Colina, L. Alonso-Llanes, E. Martínez, A. J. Batista-Leyva, C. Clément,. C. Fieldner, R. Toussaint y E. Altshuler "Lock-in accelerometry to follow sink dynamics in shaken granular matter". *Rev. Sci. Instrum*. **85:** 126101 (2014)

[18] E. Altshuler, H. Torres, A. González-Pita, G. Sánchez-Colina, C. Pérez-Penichet, S.Waitukaitis and R. C. Hidalgo "Settling into dry granular media in different gravities". *Geophys. Res. Lett.* **41**: 3032 (2014)

[19] G. Viera-López, A. Serrano-Muñoz, J. Amigó-Vera, O. Cruzata and E. Altshuler "Planetary gravities made simple: Sample test of a Mars rover wheel" *Rev. Sci. Instrum*. **88**: 086107 (2017)

[20] J. Amigó-Vera, G. Viera-López, A. Serrano-Muñoz and E. Altshuler "Measuring the performance of a rover wheel in Martian gravity". Rev. Cubana Fis. 36, 46 (2019).

[21] E. Altshuler, O. Ramos, Y. Núñez, J.Fernández, A. J. Batista-Leyva y C. Noda "Symmetry breaking in escaping ants". *Am. Nat.* **166**: 643 (2005)

[22] F. Tejera, A. Reyes y E. Altshuler "Uninformed sacrifice: evidence against long-range alarm transmission in foraging ants exposed to localized abduction". *Eur. Phys. J. – Special topics* **225**: 665 (2016)

[23] C. Noda, J. Fernández, C. Pérez-Penichet, y E. Altshuler, "Measuring activity in ant colonies" *Rev. Sci. Inst.* **77**: 126102 (2006)

[24] S. Nicolis, J. Fernández, C. Pérez-Penichet, C. Noda, F. Tejera, O. Ramos, D. J. T. Sumper y E. Altshuler "Foraging at the Edge of Chaos: Internal Clock versus External Forcing" *Phys. Rev. Lett.* **110**: 268104 (2013)

[25] E. Altshuler, M. Pastor, A. Garcimartín, I. Zuriguel y D. Maza "Vibrot, a Simple Device for the Conversion of Vibration into Rotation Mediated by Friction: Preliminary Evaluation" *PlosOne* **8**: e67838 (2013)

[26] H. Torres, V. M. Freixas and D. Pérez-Adán "The newtonian mechanics of a vibrot". *Rev. Cubana Fís*. **33**: 39 (2016) (acceso libre en www.revistacubanadefisica.org).

[27] Ch. Sholz y T. P. Poeschel "Actively rotating particles manufactured by rapid prototyping". *Rev. Cubana Fís*. **33**: 37 (2016) (acceso libre en www.revistacubanadefisica.org).

[28] Ch. Scholz, S. D'Silva and T. Poeschel "Ratcheting and tumbling motion of Vibrots". *New. J. Phys.* **18**: 123001 (2016)